\documentclass{article}
\usepackage[T1]{fontenc}
\usepackage[latin1]{inputenc}
\usepackage{graphics}

\def \Sigma{A}

\makeatletter

\usepackage[T1]{fontenc}
\usepackage[latin1]{inputenc}
\usepackage{graphics}

\makeatletter

\usepackage[T1]{fontenc}
\usepackage[latin1]{inputenc}
\usepackage{graphics}

\makeatletter

\usepackage[T1]{fontenc}
\usepackage[latin1]{inputenc}
\usepackage{graphics}
\usepackage{graphicx}
\usepackage{latexsym}
\usepackage{theorem}

\usepackage{epsfig}
\usepackage{graphicx}
\usepackage{color}
\usepackage{amssymb}
\usepackage{graphicx}
\usepackage{latexsym}
\usepackage{subfigure}
\usepackage{simplemargins}
\usepackage{bbm}

\newcommand{\ket}[1]{\ensuremath{\vert#1\rangle}}
\newcommand{\bra}[1]{\ensuremath{\langle#1\vert}}
\newcommand{\ketbra}[1]{\ket{#1}\bra{#1}}
\newcommand{\braket}[1]{\bra{#1}\ket{#1}}

\def\ensemble{{\cal E}}
\def\mixture{\ensuremath{{\cal E}} }
\def\mixturedef{\ensuremath{{\cal E}=\{(p_i,\ket{\psi_i})\}} }

\def\nin{\notin}
\def\[{\begin{equation}}
\def\]{\end{equation}}

\def\lmax{\ensuremath{l_{\max}}}

\def\oP{\overline{P}}

\def\be {\begin{eqnarray}}
\def\ee {\end{eqnarray}}
\def\bc {\begin{center}}
\def\ec {\end{center}}

\newcommand{\txt}[1]{\textrm{#1}}
\def\lbar{\overline{l}}

\newtheorem{theo}[section]{Theorem}
\newtheorem{lem}[section]{Theorem}
\newtheorem{defin}[section]{Definition}
\newtheorem{ex}[section]{Example}

\newenvironment{proof}[1][Proof]{\begin{trivlist}
\item[\hskip \labelsep {\bfseries #1}]}{\end{trivlist}}

\setallmargins{1.5in}


\makeatother
\makeatother

\begin{document}

\title{Lossless Quantum Compression}
\markboth{Caroline Rogers and Rajagopal Nagarajan}
{Lossless Quantum Compression}


\author{Caroline Rogers and Rajagopal Nagarajan}



 \maketitle


\abstract{
We describe lossless quantum compression of unknown mixtures (of non-orthogonal states)
and give an expression of the optimal rate of compression.
}


\section{Introduction}

The aim of lossless quantum compression is to compress a mixture (of possibly non-orthogonal
states) exactly and without error. It was previously thought \cite{bra00,koa02,sch01,bos02}
that compressing a mixture \mixturedef is impossible when the value $i$ of the state \ket{\psi_i}
to be compressed is unknown.
If $\mixture$ is compressed using a variable length code, then
\ket{\psi_i} might be in an unknown superposition of different lengths in which case the number
of qubits that the compressor should send to the decompressor cannot be determined.

If lossless quantum compression was impossible, this would indicate another profound difference
between classical and quantum information. If a fault tolerant implementation of quantum computation
was found, then even if a mixture contained large amounts of redundancy, it could not be compressed
without introducing errors. If lossless quantum compression was impossible then losing information
would be an inherent feature of efficient quantum computations involving communication.

In this paper, we show that lossless quantum compression is possible. In an ``always open"
model of communication, the decision ``how many qubits to transmit" does not have to be taken. We show
how to find the optimal rate of compression by looking at the probability that a state
lies in a particular Hilbert space. This gives the optimal rate of compression of both known
and unknown mixtures. Lossless quantum compression of unknown states is useful when the use of qubits
has some cost. One example of a cost is the probability of decoherence when a mixture
is passed through a noisy channel which disturbs each qubit independently with some probability. 
If the mixture is losslessly compressed, then the number
of dimensions in which it lies is minimised, hence the probability that it is disturbed
is minimised.

\section{Synopsis}

This paper is organised as follows. First we describe the background, including
the previous work on lossless quantum compression and some definitions, 
the arguments why lossless quantum compression is impossible and an asynchronous
model of quantum computation. Next we describe a model of communication
in which lossless quantum compression can take place.  We then prove the optimal rate of compression
and show how it can be used to protect a mixture from being disturbed in the presence of noise. We conclude
with ideas for future work.

\section{Background}

Lossless classical compression is an everyday application for
compressing files so that they can be stored more compactly on a hard drive
or sent more efficiently over a channel such as the internet. Lossless compression
can be used when lossy compression can not, for example, in real-time applications where
large blocks of data are unavailable. Classical lossless
compression is also useful theoretically, for example, it gives the
relative entropy between two systems $X$ and $Y$ a simple interpretation as the 
additional expected number of bits used when $X$ is compressed using the optimal compression code for 
$Y$ (than if $X$ had been compressed using the optimal compression code for $X$)

The aim of lossless quantum compression is to compress a mixture
$\mixture=\{(p_i,\ket{\psi_i})\}$ of quantum states using a variable
length quantum code so that the original mixture can be retrieved
exactly and without error. When the \ket{\psi_i}'s are orthogonal,
this is equivalent to lossless classical compression (since we can
rotate \ket{\psi_i} round to \ket{i} where \ket{i} is in the
computational basis). The challenge is therefore to encode \mixture
when the \ket{\psi_i}'s are non-orthogonal and the code words might have indeterminate lengths.

\subsection{Indeterminate Length Strings}

If we use a fixed length code to losslessly encode a mixture of states,
we do not gain any compression. Suppose we use a variable length code represented by a
unitary operation $C$ to encode $\ket{0}$ as $C\ket{0}=\ket{00}$
and \ket{1} as $C\ket{1}=\ket{111}$. Then $(\ket{0}+\ket{1})/\sqrt{2}$ is encoded as
\[C\left(\frac{\ket{0}+\ket{1}}{\sqrt{2}}\right)=\frac{\ket{00}+\ket{111}}{\sqrt{2}}\]
which does not have a determinate length. It is thus called an indeterminate length string.
\begin{defin}[Indeterminate Length String]
$\ket{\psi}=\sum_i \alpha_i \ket{i}$ is an indeterminate length quantum string if
there exists $i$ and $j$ with $|\alpha_i|>0$ and $|\alpha_j|>0$  and $l(i)\neq l(j)$.
\end{defin}
Determinate length strings of length $n$ exist in the Hilbert space $H^{\otimes n}$.
Indeterminate length strings exist in the Fock space
\[H^{\oplus}=\bigoplus_{n}^{\infty} H^{\otimes n} \]
Bostr\"{o}m and Felbinger \cite{bos02} defined two ways to quantify the lengths
of indeterminate length strings.
\begin{defin}[Lengths of Indeterminate Length Strings]
The base length $L$ of an indeterminate length string is the length of the longest part of its superposition 
\[L\left(\sum_i \alpha_i \ket{i}\right) = \max_{|\alpha_i|>0} l(i)\] The average length 
$\lbar$ of an indeterminate length quantum string is the average length of its superposition 
\[\lbar\left(\sum_i \alpha_i \ket{i}\right) = \sum_i |\alpha_i|^2 l(i)\]
\end{defin}
If we observe the length of a quantum string, then $\lbar$ gives us the expected length we observe
and $L$ gives us the maximum length we can observe. Given an indeterminate length string \ket{\psi},
neither its average length nor its base length can be observed without disturbing it.

\subsection{Can Indeterminate Length Strings be Used for Coding?}

Various papers \cite{bra00,koa02,sch01,bos02} have described
problems in using indeterminate length strings for lossless data
compression. Braunstein {\it et al} \cite{bra00} pointed out three
difficulties of data compression with indeterminate length strings.
The first is that if the indeterminate length strings are unknown to
both the sender and the receiver, then how can the time that
different computational paths take be synchronised when computations
are performed on the strings. The second difficulty is that if a
mixture of indeterminate length strings are transmitted at a fixed
speed, then the recipient can never be sure when a message has
arrived and the strings can be decompressed. The third difficulty is
that if the data compression performed by a read/write head (like a
Turing machine), then after the data compression, the head location
of the sender is entangled with the ``lengths" of the indeterminate
length string which represents the compressed data.

Koashi and Nobuyuki \cite{koa02} argued that it is impossible to faithfully encode a mixture of
non-orthogonal quantum strings. They modelled lossless data compression as taking
place on a hard disc with a maximum memory size of $N$ qubits. A compressed state
on the hard disc would be an unknown indeterminate length quantum string with base length $L$,
in which case, only the remaining $N-L$ qubits would be usable by other applications without disturbing
the compressed state. However the base length $L$ is not an observable, thus the other applications
cannot determine how many qubits are available. Thus the remaining $N-L$ qubits are not available
for other applications to use unless $L$ is the length of the longest code word.

Schumacher and Westmoreland \cite{sch01} envisaged that
indeterminate length quantum strings would be padded with zero's to
create determinate length strings. Each code word of a variable
length code would be padded with zero's so that each code word had
the same length. They modelled the data compression as taking place
between two parties Alice and Bob in which Alice sends Bob only the
original strings (with the zero-padding removed) leaving Alice with
a number of zero's depending the length of the string she sent. If
she sends Bob an indeterminate length string, then after the
transmission Alice and Bob are entangled. This is illustrated with an example.
\begin{ex}[Lossless Quantum Compression with Zero-padding]
If $X$ is a prefix free set given by $X=\{\ket{0},\ket{10},\ket{110},\ket{111}\}$
then after zero-padding, $X$ is transformed into the set $X'=\{\ket{000},\ket{100},\ket{110},\ket{111}\}$.
Alice starts off with a zero-padded string Hilbert space spanned by $X'$.
If she starts off with the state \ket{000} or \ket{100}, then she sends Bob \ket{0} or \ket{10}
respectively and she is left with the state \ket{0} or \ket{00} respectively. If she starts off with the state
$(\ket{000}+\ket{100})/\sqrt{2}$ then she sends $(\ket{0}+\ket{10})/\sqrt{2}$ and is left
with the state $(\ket{00}+\ket{0})/\sqrt{2}$. By measuring the state $(\ket{00}+\ket{0})/\sqrt{2}$,
she can collapse the state $(\ket{0}+\ket{10})/\sqrt{2}$ which Bob has received.
\end{ex}
Since Alice can disturb the state she has sent to Bob, this
scheme is an unsuitable model of lossless quantum data compression.

Bostr\"{o}m and Felbinger \cite{bos02} pointed out that quantum prefix strings are not useful.
Classical prefix strings carry their own length information,
however the length information indeterminate length prefix strings is unobservable without
disturbing the string. They also considered zero-padding and said that in such a scheme
an unknown indeterminate length quantum string could not be transmitted because the number
of zeros to remove before the string is transmitted cannot be determined without disturbing it.

\subsection{Properties of Indeterminate Length Strings}

Schumacher and Westmoreland \cite{sch01} investigated the general properties of
indeterminate length strings. An indeterminate length string can be
padded with zeroes so that its length becomes an observable.
\begin{defin}[Zero Extended Form]
If $\ket{\psi}=\sum_{i<2^{l_{\max}}} \alpha_i \ket{i}$ is a quantum string in
a register of $\lmax$ qubits, then its zero-extended form is:
\[
\ket{\psi_{zef}} =\sum_{i<2^{\lmax}} \alpha_i \ket{i 0^{\otimes \lmax - l(i)}}
\]
\end{defin}
Given a sequence of $N$ strings, it is useful to be able to condense them so that
the strings are packed together at the beginning of the string and the zero-padding
all lies at the end of the sequence.
\begin{defin}[Condensable Strings]
A set of strings $\xi$ is condensable if for any $N$, there exists a unitary operation
$U$ such that:
\[U(\ket{\psi^{1}_{zef}}\otimes \ldots \otimes \ket{\psi^{N}_{zef}}) = (\ket{\psi^{1}}\otimes \ldots \otimes \ket{\psi^{N}})_{zef}\]
\end{defin}
It is easy to see that superpositions of classical prefix free strings are condensable.
Prefix strings were defined more generally.
\begin{defin}[Zero-Padded Prefix Free Strings]
Suppose \ket{\psi^{1}} and \ket{\psi^{2}} are quantum strings with $L(\ket{\psi^1})>L(\ket{\psi^2})$
and that they are in a register of $\lmax$ qubits.
The first $l_1$ qubits of \ket{\psi^{2}_{zef}} may
be in a mixed state, described by the density operator
\[ \rho^{1 \ldots l_1}_2 = tr_{l_1+1 \ldots \lmax} (\ket{\psi^{2}_{zef}})
\]
\ket{\psi^1} and \ket{\psi^2} are prefix free if:
\[
\bra{\psi_{1\ldots l_1}^1} \rho^{1 \ldots l_1}_2 \ket{\psi_{1\ldots l_1}^1}
\]
where \ket{\psi_{1\ldots l_1}^1}  denotes the first $l_1$ qubits of \ket{\psi_1}'s zero-extended form.
\end{defin}

\subsection{From Lossless Coding to Lossy Coding}

Schumacher and Westmoreland \cite{sch01} demonstrated
that by projecting onto
approximately $n(S(\ensemble)+\delta)$ qubits, if a mixture $\ensemble^{\otimes n}$
is encoded with a variable length condensable code, a fixed length lossy code can be obtained \cite{chu00}.
If $\ensemble$ is a mixture with density operator $\rho$, where $\rho$'s spectral
decomposition is:
\[\rho = -\sum_i p_i \ketbra{i}\]
Then \mixture can be encoded by encoding each \ket{i} as a prefix free string
of length $\lceil-\log(p_i)\rceil$ with zero-padding.
$\rho^{\otimes n}$ can be encoded in the same
fashion. Almost every string in the typical subspace of $\rho^{\otimes n}$ has probability
arbitrarily close to $2^{-nS(\rho)}$ as $n$ grows large. Thus almost every string in the typical subspace of $\rho$
is encoded as a string of length arbitrarily close to $nS(\rho)$. By projecting onto $n(S(\rho)+\delta)$ qubits, we project onto the encoded
typical subspace of $\rho$. We can decode the typical subspace to obtain the original mixture
\mixture with arbitrarily high (but not perfect) probability and fidelity.
Thus we can use a variable length code to design a lossy code.

From this encoding, we can see that the average lengths of  condensable codes obey Kraft's inequality (if they did not, then
we could lossily compress a mixture to less than its von Neumann entropy). Since the base length of a string is
bounded below its average length so Kraft's inequality also holds for the base lengths.
\begin{lem}[Kraft's Inequality for Condensable Strings]
If $\xi$ is a set of orthogonal condensable strings then
\[
\sum_{\ket{\psi}\in \xi} 2^{-L(\ket{\psi})} \leq \sum_{\ket{\psi}\in \xi} 2^{-\lbar(\ket{\psi})} \leq 1 
\]
\end{lem}

\subsection{Coding Information With Classical Side Channels}

Bostr\"{o}m and Felbinger \cite{bos02} gave a scheme for lossless
quantum compression using classical side channels. If \mixturedef is
the mixture to be compressed, then they assume that the value of $i$
is known to the compressor, Alice. If she encodes \mixture using a
unitary operation $U$, then she sends the base length of the
compressed string to Bob, the decompressor, through a classical side
channel. She then sends $L(C(\ket{\psi_i}))$ qubits of
\ket{\psi_i}'s zero-extended form to Bob. Since the length of the
encoded string is encoded classically, it is not necessary to use a prefix free
code to encode the quantum part --- thus $C$ is unitary but not necessarily condensable.

Rallan and Vedral \cite{ral02} gave another scheme for lossless
quantum compression with classical side channels which does not use
zero extended forms. They envisaged that the compressed state would
be represented by photons --- thus using a tertiary alphabet
$\{\ket{0},\ket{1},\ket{\epsilon}\}$ where \ket{\epsilon} denotes
the absence of a photon and marks the end of the string. They
assumed that the Alice has $n$ copies of a mixture \mixture which
she would like to send to Bob. In this scheme, Alice only sends Bob
the value of $n$. This
scheme has a nice physical interpretation.

\subsection{Physical Interpretations of Indeterminate Length Strings}

Bostr\"{o}m and Felbinger \cite{bos02} pointed out that variable length
quantum strings can be realised in a quantum system whose particle
number is not conserved. Rallan and Vedral \cite{ral02} described in
detail an example system where the average length of a string can be
interpreted as its energy. A Hilbert space $H^{\otimes n}$ can be
realised by a sequence of photons $\ket{\phi_1}\otimes \ldots
\otimes \ket{\phi_n}$ in which \ket{\phi_i} represents exactly one
photon with frequency $\omega_i$. The value of the qubit
\ket{\phi_i} is realised by the polarisation of its photon, either
horizontal \ket{0} or vertical \ket{1}. The absence of a photon can
be represented by \ket{\epsilon} which is orthogonal to \ket{0} and
\ket{1}. We obtain indeterminate length strings by allowing the
number of photons to exist in superposition. The frequency of each
photon \ket{\phi_i} is chosen to be equal so that $\omega_i \approx
\omega$ for some value $\omega$. The energy in a superposition of photons
is the average energy required to either create or destroy that
superposition ($\hbar \omega$ per photon of frequency $\omega$).
Thus the energy of an indeterminate length string of photons
\ket{\phi} is proportional to its average length and is given by
$\hbar \omega \lbar(\ket{\phi})$. In this interpretation, lossy data
compression can be interpreted as the average energy required
destroy a mixture \mixture since destroying a mixture is equivalent
to sending it to the environment (which is another recipient).

\subsection{Asynchronous Model of Quantum Computation}

If quantum computers are to be used to solve classical problems
efficiently, then it seems reasonable to demand that all the paths
of a quantum Turing machine \cite{deu85,ber97} halt simaltaneously.
However this demand raises various issues.
If two strings with different halting times are input in superposition,
then the resulting computation halts at a superposition of different times \cite{mye97}.
It is uncomputable to say whether an arbitrarily constructed quantum Turing machine
halts at a deterministic time \cite{miy02}. A quantum Turing machine that
halts is not unitary since it cannot be reversed after the computation has halted \cite{kie98}.

These issues were resolved by Linden and Popescu \cite{lin98} who described
a quantum Turing machine augmented with an ancillary system in which
computations take place after the Turing machine has ``halted". 
The ancillary system can record the time since computation 
began so that the output is disentangled from the time at which it ``halts".
Thus there is a well-defined model of quantum computation in which computation
paths halt at different times.

\section{Communication Model for Lossless Quantum Compression}

We have discussed the arguments why lossless quantum compression is impossible \cite{bra00,koa02,sch01,bos02}.
Now we describe how lossless quantum compression of unknown mixtures is possible
by taking an appropriate model of communication. For quantum compression, there are two cases,
the mixture to be compressed can be known or unknown. We show that the same holds for classical
compression depending whether the decision on what data is to be compressed is made before
or after the data has been read.

\subsection{Lossless Quantum Compression of Known Mixtures}

Bostr\"{o}m and Felbinger \cite{bos02} gave a scheme for losslessly
compressing a mixture \mixturedef in which the value $i$ of the
state to be compressed is known to the compressor, Alice. Thus she
can deduce the base length $L(\ket{\psi_i})$ of the string to be
compressed and transmit this many qubits to Bob through a quantum
channel and send $L(\ket{\psi_i})$ through a classical channel.
Since Bob knows the value of $L(\ket{\psi_i})$, the compressed
quantum states are not necessarily prefix free (i.e. condensable).

However the string
$\ket{\Psi}=L(\ket{\psi_i})\otimes \ket{\psi_i}$ which represents the classical and quantum parts
together is not prefix free. An example of a prefix free encoding is
\[\ket{\Psi'}=1^{\lceil \log (L(\ket{\psi_i})) \rceil}0L(\ket{\psi_i})\otimes \ket{\psi_i}\]
where $1^{\lceil \log (L(\ket{\psi_i})) \rceil}0L(\ket{\psi_i})$ is
sent through the classical side channel. (To find the length of
\ket{\Psi'}, we find the length of the first contiguous sequence of
$1$'s followed by a $0$. Then we read the next $\lceil \log
(L(\ket{\psi_i})) \rceil$ to find the length of $\ket{\psi_i}$.)
Thus with a slight modification to Bostr\"{o}m and Felbinger's
scheme, we have a prefix free encoding of known quantum mixtures
where the length of the encoded data can be read from the classical
part. However, an important question remains open: ``What is the
rate of compression?". Since this scheme is based on one-one coding
rather than prefix free encoding, the analysis of the compression
rate is more tricky though there are known bounds between the two
compression rates classically \cite{blundo,li97,leung}. Instead, we search
for a prefix free quantum encoding and analyse its compression rate.
This will also enable us to compress unknown mixtures. But first, we
resolve the issues of compressing with unknown indeterminate length
strings.

\subsection{How Much Memory Is Free?}

Koashi and Nobuyuki \cite{koa02} modelled data compression as taking place on a
computer where only $N$ qubits of memory are available. Let $C$ be a
classical prefix code for a random variable $X$. A na\"{i}ve guess
is that, on average, we can losslessly compress $N/H(X)$ copies of
$X$ into a memory of $N$ bits. Let $l_{\max}$ be the length of $C$'s
longest code word and let $n$ be the number of copies of $X$ we
compress into the memory. Then, in the worst case, $X^n$ compresses
to $n l_{\max}$ bits. If $n$ is chosen to be greater than
$N/l_{\max}$, then with some small probability of error, the
compression fails. Thus we can compress $X^n$ losslessly only when
$n\leq N/l_{\max}$. (If we want to perform a computation on the
remaining bits, then if we decide the computation in advance of the
compression, there are only $N-n\lmax$ bits available.)

Both classical and quantum lossless compression can be modelled in two ways, depending
on whether the states to be compressed are known or unknown.
\begin{description}
\item[Ad Hoc ``Known" Compression] Classically, once the value of $X$ is known, it can be deduced
how much memory is free --- the compression takes place {\it ad hoc} in
that the decision whether there is enough space free to compress is
decided by examining the memory at the time of compression. We find
an analogous quantum situation when the value of $\ensemble$ is
known. In which case we can use Bostr\"{o}m and Felbinger's scheme \cite{bos02} to record classically the amount of the memory which
has been used so that the amount of free space available is known.
\item[Reversible ``Unknown" Compression] The classical analogy of compressing a mixture when its
value is unknown is describing the compression of a random variable $X$ before its value is known.
If Alice decides to compress $X$ and then to perform a computation in the free memory, she has to
assume the worst case compression rate in order for the combined compression--computation to be reversible.
The same situation arises in the quantum case when the value of the mixture being compressed is unknown.
\end{description}
In the unknown quantum case, only some branches of computation may fail through lack of memory, in which case the compressor cannot be
sure whether the computation has succeeded at a later time.
In the unknown classical reversible case, the compressor can measure at a later time
with certainty whether a computation has failed through lack of memory.
However, in advance of the compression, the compressor does not know whether the
computation has failed.

\subsection{How Many Qubits to Transmit?}

If Alice has an unknown indeterminate length quantum string,
how can she decide how many qubits to transmit to Bob? Consider the following example.
\begin{ex}[Open and Closed Channels]
Alice and Bob have mobile phones which they leave switched on all the time.
Alice says to Bob that she will phone him at 7pm if she is available to have dinner.
If Alice does not phone Bob at 7pm, Bob can deduce that Alice is not available to have dinner.
Whenever the phones are switched on, the channel is open and information is being exchanged.
\end{ex}
Thus if there is a channel between Alice and Bob, then Alice and Bob are always communicating \cite{bowen}.
We can represent an ``open--closed" channel with a tertiary alphabet $\{0,1,\epsilon\}$
where $\epsilon$ denotes ``no communication". If Alice sends Bob a string in such a channel,
then there is no need to use a prefix code since the closure of the channel marks the end of the sequence.

\begin{figure}[t]
\begin{center}
\resizebox{9cm}{!}{\includegraphics{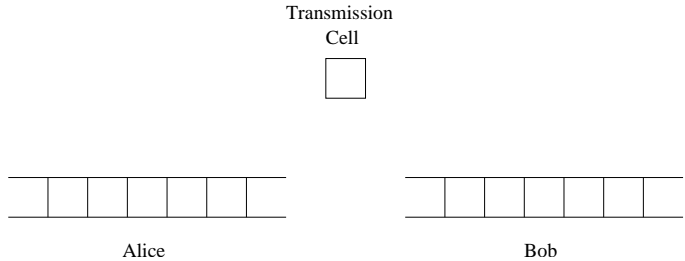}}
\end{center}
\caption{A channel between Alice and Bob. At each time step,
Alice can read or write from the transmission cell, then Bob can read and write from the transmission cell.
Using this channel, Alice can reversibly send Bob a string. She condenses the string
in the initial part of her memory and pads it with zeroes. The transmission cell and Bob's memory
are initially prepared as zeroes. To send a message to Bob, at step $i$, Alice swaps the $i$th (qu)bit in her memory
with the value in the transmission cell, then Bob swaps the value in the transmission cell with the $i$th bit in his memory.
A string of base length $l$ is transmitted in $l$ steps. \label{communication}}
\end{figure}
A model of an always open channel \cite{bowen,nielsen} is shown in Fig. \ref{communication}.
We do not require an $\epsilon$ no communication character since the channel is always open.
If Alice wants to send Bob
a sequence of condensable indeterminate length strings, she condenses them and sends
the qubits one by one. By assuming that Bob's memory is also padded with zeroes,
we avoid the entanglement issues described by Schumacher and Westmoreland \cite{sch01}.
As the transmission is taking place, Alice is free to add append additional
unknown condensable quantum strings onto those being transmitted. However, neither Alice or Bob can measure
whether a string has been transmitted.

\subsection{When Can Bob Decompress?}

Suppose Bob wants to decompress the strings as they arrive? How can he decide when to begin decompression?
In the standard model of quantum computation \cite{nielsen,vlatkobook}, computations begin and end at determinate times.
However as Linden and Popescu showed \cite{lin98}, there is a well-defined model of quantum computation
where computations can begin and end at superpositions of different times. Like when compressing onto a finite memory,
we have two cases, whether the mixture to be compressed is known or unknown which correspond to the
two classical cases whether the protocol is decided in advance or {\it ad hoc}. If the mixture is unknown and Bob
wants to perform a measurement on the decompressed state, then he waits for the maximum possible time of transmission
and decompression before making the measurement. Similarly, if a random variable $X$ is classically compressed, then
if the time of a measurement is decided before the value of $X$ is known, then the measurement may fail.

\subsection{Lossless Quantum Compression of Unknown States}

We have described two situations for lossless quantum compression of a mixture
\mixturedef of non-orthogonal states. When the mixture is known
(i.e. the value of $i$ is known to the compressor), Bostr\"{o}m and
Felbinger's scheme \cite{bos02} can be used and the lengths of the
encoded data are an observable. When the mixture is unknown, the
mixture can be compressed and transmitted using a condensable code
as shown in Fig. \ref{communication}. The expected average length
$E(\lbar(\mixture))$ of the optimal code is known to be
approximately the von Neumann entropy of $\mixture$, but in order to
keep a mixture of indeterminate length strings intact, the number of
qubits of each string that need to be kept intact is its base
length. However, in either the known or unknown case, the optimal
rate of compression in terms of the base lengths is still open. We will show, by assigning
probabilities to Hilbert spaces according to the probability that
string lies in a space, that the optimal rate of compression can be
found by finding the most probable Hilbert spaces first.

\section{Prefix Free Strings}

Lossless quantum compression makes use of prefix free strings. Schumacher and Westmoreland \cite{sch01}
defined prefix free quantum strings in terms of their zero-extended forms using the trace operator.
It is simpler just to directly generalise the classical definition.

\begin{defin}[Prefix Free Quantum Strings]
A string \ket{\phi} is the prefix of a string \ket{\psi} if there exists a string \ket{\chi}
with $|\langle \epsilon \ket{\chi}|=0$ such that
\[|\langle \phi \chi | \psi \rangle|>0\]

A set $\xi$ of quantum strings is prefix free if any two (not necessarily distinct) strings in $\xi$
are prefix free.
\end{defin}

Unlike deterministic classical strings, deterministic quantum strings can be prefixed by themselves.
For example,
\[\ket{\psi}=\frac{\ket{0}+\ket{00}}{\sqrt{2}}\]
is a prefix of itself.
In classical information theory, the empty string $\epsilon$ is not prefix free since
it multiple copies of $\epsilon$ are not uniquely decipherable. Superpositions of the empty string
\ket{\epsilon} are self-prefix since if $\ket{\psi}=\alpha \ket{\epsilon}+\beta \ket{\phi}$ then
$|\langle \psi \ket{\psi \phi}| =  |\alpha \beta \braket{\phi}|$.

As did Bostr\"{o}m and Felbinger \cite{bos02}, we can define Hilbert
spaces with prefix free bases.
\begin{defin}[Prefix Free Hilbert Space]
A Hilbert space $H$ is prefix free if it has a basis of prefix free strings.
\end{defin}
 We check such Hilbert spaces are well-defined by showing that any orthogonal basis
 of a prefix free Hilbert space is prefix free.

\begin{lem}[Prefix Hilbert Spaces are Well-defined]
If $H$ is a prefix Hilbert space which is the span of a sequence of prefix free strings $\xi_1$, $\ldots$, $\xi_n$,
then any orthogonal basis for $H$ is prefix free.
\end{lem}
\begin{proof}
To show this holds, we show that any string \ket{\psi} in $H$ is not a prefix of itself and that
any two orthogonal strings \ket{\phi} and \ket{\psi} in $H$ are prefix free.

Let \ket{\psi} be any string in $H$. Then \ket{\psi} can be expressed as
\[\ket{\psi}= \sum_i \alpha_i \ket{\xi_i} \]
Let $\ket{\chi}$ be any quantum string with $|\langle\epsilon | \chi\rangle|=0$. Then
\[
| \langle \psi | \psi \chi\rangle | = \left| \sum_{i,j} \alpha_{i}^* \alpha_j \bra{\xi_i}\xi_j \chi\rangle\right|
\]
Since the $\ket{\xi_i}$'s form a prefix free set, $|\bra{\xi_i}\xi_j \chi\rangle|=0$ for all $i$ and $j$,
hence
\[
| \langle \psi | \psi \chi\rangle | =0
\]
and \ket{\psi} is not a prefix of itself.

Now we show that any two orthogonal strings \ket{\phi} and \ket{\psi} in $H$ are prefix free.
Again, let $\ket{\chi}$ with $|\langle \epsilon | \chi \rangle|=0.$
We can express \ket{\phi} and \ket{\psi} as
\be
\ket{\psi} &=& \sum_i \alpha_i \ket{\xi_i} \\
\ket{\phi} &=& \sum_j \beta_j \ket{\xi_j}
\ee
Then using the prefix free property of the \ket{\xi_i}'s,
\be
|\langle \phi \chi | \psi \rangle|
&=& \left| \sum_{i,j} \beta^{*}_j \alpha_{i} \langle \xi_j \chi | \xi_i \rangle     \right| \\
&=& 0
\ee
which completes the proof.
\end{proof}
Prefix free Hilbert spaces can be placed side by side so that their elements are condensable.
\begin{lem}
A set of strings in a prefix free Hilbert space is condensable.
\end{lem}
\begin{proof}
Let $\{\xi_i\}_i$ be a basis for a prefix free Hilbert space $H$.
Let $\lmax$ be the length of the longest base length of a string in $H$ (i.e. the size of the
register). For each $\ket{\psi}\in H$, let $\ket{\psi^{zef}}$ be its zero-extended form  (so that
\ket{\psi} is padded out with zeros to form a string of determinate length $\lmax$). 
Given an integer $n$, we can design a unitary operation $U_n$ on the basis vectors $\xi_i$ so that
\[U( \ket{\xi^{zef}_1} \ldots \ket{\xi^{zef}_n})=(\ket{\xi_1}\otimes \ldots \ket{\xi_n})^{zef}\]
$U$ is reversible and hence unitary. $U$ condenses any set of strings $\xi$
drawn from $H$.
\end{proof}

Since any set of orthogonal prefix free strings are condensable, their average and base lengths
obey Kraft's inequality  \cite{sch01}. If Alice wants to send Bob a sequence of strings from prefix free Hilbert spaces,
she can condense them and send them as shown in Fig \ref{communication}.

\section{Lossless Quantum Data Compression}

The aim of lossless quantum data compression is, using as few qubits as possible,
to encode a mixture of non-orthogonal states. When the states are orthogonal, the mixture can be encoded
using determinate length strings and the
rate of compression is simply the von Neumann entropy of the mixture. 
When the states are non-orthogonal, the expected length of the encoding
is the expected base length of the compressed strings, since this is the minimum number of
qubits that must be left intact for the mixture to be retrievable exactly and without error.

\begin{defin}[Lossless Quantum Code]
Let $\ensemble=\{p_i, \ket{\psi_i}\}_i$ be a mixture of quantum states in a Hilbert space $H$.
A lossless code $C$ is a unitary operation from $H$ to a prefix free Hilbert space $H'$.
If $B$ is an orthogonal basis for $H$ then $C(B)$ is a set of code words for $C$.

The expected length of compression of $C$ is:
\[
E(L(C(\ensemble))) = \sum_i p_i L(C(\ket{\psi_i}))
\]

$C$ is optimal if for any other code $C'$,
\[E(L(C(\ensemble)) \leq E(L(C'(\ensemble)))\]
\end{defin}

\begin{figure}[t]
\begin{center}
\resizebox{9cm}{!}{\input{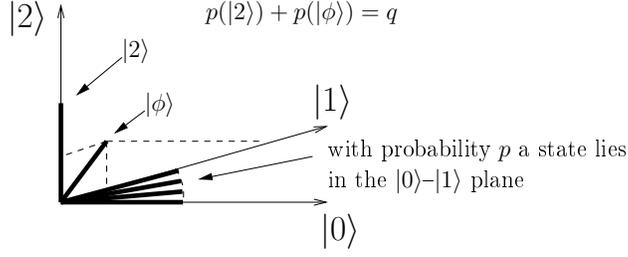}} \caption{
An example of lossless quantum compression. With high probability, say $p$, a state lies
in the \ket{0}--\ket{1} plane. We say that the probability of the plane is $p$. 
Since the plane is spanned by two vectors, \ket{0} and \ket{1}, we say that the average
probability of the plane is $p/2$ and encode the plane as strings of length $-\log(p/2)$ by encoding
\ket{0} and \ket{1} as strings each of length $-\log(p/2)$. There are two non-orthogonal strings \ket{2} and \ket{\phi} which are outside
the \ket{0}--\ket{1} plane, suppose the probability of these two strings sums to $q<p/2$. Then we can
encode $\ket{2}$ as a string of length $-\log(q)$, in which case, the string \ket{2} is encoded as a string
of determinate length \ket{2}. The other string \ket{\phi} is encoded as a string in a superposition of lengths
$-\log(p/2)$ and $-\log(q)$, so that its base length is $-\log(q)$. Since we encode in this way, we say that the probability of \ket{2}
with respect to the \ket{0}--\ket{1} plane is $q$.
\label{LosslessQuantumCompression}}
\end{center}
\end{figure}

An example of lossless quantum compression is shown in Fig. \ref{LosslessQuantumCompression}. 
In analysing lossless codes, it is convenient to define probability
in terms of subspaces. The idea is to encode small subspaces with
high probability using short codes. We define the probability $P(X)$
of a subspace $X$ to be the total probability of all strings lying
completely within $X$. We can share the probability of $X$ equally
between its basis vectors and define the average probability
$\overline{P}$ of $X$ to be $\overline{P}(X)/\dim(X)$. If a subspace
$Y$ has a  large average probability, then it can be encoded
with short strings. The average probability of another subspace $X$
might be very small, but if there is a reasonably large probability
that strings lie in the space $X \oplus Y$, then we can encode $X$
with reasonably short strings so that the strings that lie both in
$X$ and $Y$ are encoded with a reasonably small base length. We
define the average probability $\overline{P}$ of a subspace $X$ with
respect to a subspace $Y$ to be the probability that a string lies
partially in $X$ given that it lies completely within $X \oplus Y$
divided by the dimension of $X$.
\begin{defin}[Subspace Probabilities]
Let $\ensemble = \{(p_i,\ket{\psi_i})\}$ be a mixture of quantum states in the Hilbert space $H$.
If $X$ is a subspace of $H$, then the probability $P$ and average probability $\overline{P}$ of $X$ are
\be
P(X) &=& \sum_{\ket{\psi_i} \in X} p_i \\
\oP(X) &=& \frac{P(X)}{\dim(X)}
\ee
The probability $P(X:Y)$ of a subspace $X$ with respect to a subspace $Y$ is the sum of the strings in the space $X \oplus Y$
which are partially within $X$. The average probability $\oP(X:Y)$ of a subspace $X$
with respect to a subspace $Y$ is the sum of the strings in the space $X \oplus Y$
which are partially within $X$ divided by the dimensions of $X$.
\be
P(X:Y) &=& \sum_{\ket{\psi_i} \in {X\oplus Y} \txt{ and } \ket{\psi_i}\nin Y} p_i \\
\oP(X:Y) &=& \frac{P(X:Y)}{\dim(X)}
\ee
\end{defin}

A space $H$ might contain some subspaces which have higher average probabilities than others.
We can decompose $H$ into its subspaces by finding the largest subspace $X_1$ which has highest average probability first,
then finding the largest subspace $X_2$ which has highest average probability with respect to $X_1$, then finding
the largest subspace $X_3$ which has highest average probability with respect to $X_1$ and $X_2$ and so on.
Let $P_i$ be a projection onto the subspace $X_i$. We can define
a density operator by summating these projections where the eigenvalues for $P_i$ are given by
the average probability of $X_i$ with respect $X_{1\ldots i-1} = X_1 \oplus \ldots \oplus X_{i-1}$.

\begin{defin}[Decompositions of a Hilbert space by mixture ]
Let $\ensemble = \{(p_i,\ket{\psi_i})\}$ be a mixture of quantum states in the Hilbert space $H$.
We define subspace decomposition of $H$ as $X_1$, $X_2$, $\ldots$, $X_m$ where $X_i$ is defined as:
\be X_1 &=& X : \oP(X) > \oP(X')\txt{ } \forall X'\subseteq H \txt{ and } X'\neq X\\
X_{i+1} &=& X : \oP(X) > \oP(X' : X_{1\ldots i}) \nonumber \\
&&\forall X'\subseteq H-(X_{1\ldots m}) \txt{ and } X'\neq X
\ee
where $X_{1\ldots i}=X_1 \oplus \ldots \oplus X_i$.
Let $P_i$ be the projection onto $X_i$. Then the density operator decomposition of $H$ is the density operator $\rho$ defined as:
\[
\rho = \sum_i \oP(X_i : X_{ 1\ldots {i-1}}) P_i
\]
\end{defin}

The probability each string \ket{\psi_i} is only counted once so the density operator decomposition has trace $1$.
\[\sum_i \sum_i \oP(X_i : X_1 \oplus \ldots \oplus X_{i-1}) \dim(X_i) = 1\]
Since the subspaces are orthogonal to one another, we can use the converse of Kraft's inequality to
encode each basis vector of each $X_i$ as a prefix string of length of $\lceil\log(\oP(X_i : X_1 \oplus \ldots \oplus X_{i-1}))\rceil$.
We now show that this is the optimal encoding.

\begin{theo}[Noiseless Coding Theorem for Lossless Quantum Codes]
Let $\ensemble = \{(p_i,\ket{\psi_i})\}$ be a mixture of quantum states in the Hilbert space $H$.
Let $X_1$, $\ldots$, $X_m$ be the decomposition of $H$ with density operator
\[
\rho = \sum_i \oP(X_i : X_{1 \ldots i-1}) P_i
\]
Let $Z_l = \bigoplus_{i: 2^{l+1} < \oP(X_i:X_{1\ldots i-1}) \leq 2^l }   X_i$ ($Z_l$ is the space of strings
that are encoded as strings of length $l$).
Then there is a prefix free code $C$ such that for all $\ket{\psi}\in Z_l$,
\[L(C(\ket{\psi})) \leq l\]
where the expected length of $C$ is bounded by:
\[S(\rho) \leq E(L(C(\mixture))) \leq S(\rho)+1\]
and for any other prefix code $C'$:
\[E(L(C(\mixture))) \leq E(L(C'(\mixture))) +1\]
\end{theo}
\begin{proof}
Let $C'$ be any prefix free lossless quantum code on $H$. The proof proceeds as follows.
\begin{itemize}
\item First we show by induction that if $C'$ is a prefix code then $H$ can be divided up into orthogonal subspaces
$Z_{l}'$ which are encoded  with base length $l$.
\item Next we show that if $C'$ is optimal, then the average probability of $Z_{l}'$ with respect to $Z_{1\ldots l-1}'$
is about $2^{-l}$.
\item We show by induction that if $C'$ is optimal then $Z_{l}' \subseteq Z_{1\ldots l}$ for all $l$ which shows that for any \ket{\psi},
$L(C'(\ket{\psi}))\geq L(C(\ket{\psi}))$.
\end{itemize}

Let $Z_1$ be the set of strings $\ket{\psi}\in H$ such that $L(C'(\ket{\psi}))=1$.
Then $Z_1$ forms a subspace as any string of base length $1$ has determinate length
$1$.
Let $Z_{l+1}$ be the set of strings $\ket{\psi}\in H-(Z_{1\ldots l})$ such that $L(C'(\ket{\psi}))=l+1$.
Then assuming that $Z_1$, $\ldots$, $Z_l$ form subspaces, so does $Z_{l+1}$ since
if $\ket{\psi_1}$ and $\ket{\psi_2}$ are two strings in $Z_{l+1}$ with
$L(\alpha \ket{\psi_1}+\beta \ket{\psi_2})< l+1$ then $\alpha \ket{\psi_1}+\beta \ket{\psi_2} \in Z_k$ where $k<l+1$.
Since, by our assumption, $Z_k$ is a subspace, if $\alpha \ket{\psi_1}+\beta \ket{\psi_2} \in Z_k$ then $\ket{\psi_1}$ and \ket{\psi_2}
are not in $H-(Z_1 \oplus \ldots \oplus Z_l)$ and hence not in $Z_{l+1}$.

Now we use Shannon's noiseless coding theorem for lossless codes to show that
$\oP(Z_{l}':Z_{1\ldots l-1})\approx 2^{-l}$ if $C'$ is optimal.
The expected rate of encoding of $C'$ is:
\[
E(L(C'(\ensemble))) = \sum_l \oP(Z_{l}':Z_{1\ldots l-1}) \dim(Z_{l}') l 
\]
Since $C'$ is prefix free, we have $\sum_l \dim(Z_{l}') 2^{-l}\leq 1$. Thus according to Shannon's noiseless coding theorem for lossless
codes, $E(L(C'(\ensemble)))$ is minimal to within one qubit if $C'$ is chosen so that $Z_{l}'$ is encoded using strings of length
$\lceil -\log (\oP(Z_{l}':Z_{1\ldots l-1})) \rceil $, in other words if for each $l$,
\[
\lfloor -\log (\oP(Z_{l}':Z_{1\ldots l-1}')) \rfloor \leq l \leq \lceil -\log (\oP(Z_{l}':Z_{1\ldots l-1}')) \rceil \label{aneq}
\]
Thus the encoding of $C'$ is much like the encoding of $C$ except that the subspaces could be chosen differently.

We now assume that for each $l$:
\[l = \lfloor -\log (\oP(Z_{l}':Z_{1\ldots l-1}')) \rfloor\]
this only changes the expected length of compression of $C'$ by one qubit and show by induction
that this implies that $Z_{l}'\subseteq Z_{1 \ldots l}$. 
$Z_{1}'\subseteq Z_1$ since $Z_1$ is chosen to contain all the
subspaces of average probability at least $1/2$. Assuming that $Z_{k}' \subseteq Z_{1\ldots k}$ for all
$k\leq l$, for any subspace $X$, $P(X:Z_{1\ldots l}')\leq P(X:Z_{1\ldots l})$. If $Z_{l+1}'$ was not in
$Z_{1\ldots l+1}$, then $Z_{l+1}'$ would contain a subspace $X$ such that $P(X:Z_{1\ldots l}') \geq 1/2^{l+1}$ but
this is a contradiction.

Since for each $l$, $Z_{l}' \subseteq Z_{1\ldots l}$, if $L(C'(\ket{\psi}))=l$ then $L(C(\ket{\psi}))\leq l$.
To prove this we have assumed that
\[l = \lfloor -\log (\oP(Z_{l}':Z_{1\ldots l-1}')) \rfloor\]
However, if $C'$ is optimal from Eq. \ref{aneq}, the expected length of $C'$ could be one qubit less.
Thus $C$ is optimal to within $1$ qubit.
\end{proof}

\section{Using Variable Length Compression in Noisy Channels}

Bostr\"{o}em and Felbinger  \cite{bos02} suggested there may be a relationship between error correction
and variable length coding. The optimal code compresses highly probable subspaces so that
the expected number of dimensions in which a string lies is minimised.
Suppose an encoded mixture was sent through a noisy channel which
introduces errors each qubit independently. Then by minimising the expected number of dimensions
in which the mixture lies, the probability of its disturbance is minimised. An alternative is to variable
length code in the diagonal basis of the mixture's density operator, in which case the probability
of decoherence might be higher, but the expected probability of being able to distinguish the initial
and final states is smaller. We provide a simple example to illustrate the differences between the two
schemes.
\begin{ex}[Lossless Quantum Compression to Prevent Noise]
Let $\mixture$ be a mixture of quantum states with probabilities:
\be
P(\sqrt{(1-\delta)}\ket{0}+\sqrt{\delta}\ket{1}) &=& 1/2 \\
P(\sqrt{(1-\delta)}\ket{0}-\sqrt{\delta}\ket{1}) &=& 1/2
\ee
Then the density operator $\rho$ which represents the mixture $\mixture$ is:
\[
\rho = (1-\delta)\ketbra{0}+\delta\ketbra{1}
\]
We now consider compressing \mixture using the base length compression scheme described in the previous section
and compressing \mixture in the diagonal basis of its density operator so that it's average length is minimised.
We assume that each qubit in the encoded state is disturbed with probability $p$. 
\begin{description}
\item[Base Length Compression] If \mixture is encoded to minimise its expected base length, then the two strings
are encoded as strings of determinate length $1$. The probability that the encoded mixture is disturbed is $p$.
\item[Average Length Compression] If \mixture is encoded in the diagonal basis, then $\ket{0}$ is encoded as a string
of average length $-\log(1-\delta)$ and \ket{1} is encoded as a string of average length $-\log(\delta)$ (assuming e.g. that
there are a large number of copies so that the strings can be encoded with non-integer lengths on average). When $\delta$
is small, each state in the mixture is a superposition of a very short string with high amplitude and a very long string
with small amplitude. The probability that the whole state is disturbed is very small ($p^{-\log(1-\delta)}$) but 
with large probability $(p^{-\log(\delta)})$ the state suffers a very small disturbance.
\end{description}
\end{ex}
Base length compression minimises the probability that a state is disturbed at all whereas average length compression
minimises the probability that decompressed state can be distinguished from the original. In either case, error correction
can be used to amplify the probability that the mixture is not disturbed. When applied to the base length compression scheme,
it amplifies the probability that the mixture is left completely intact.

\section{Conclusions}

We have given a model of communication for lossless quantum compression
and shown how to find the optimal code and rate. 
We now describe avenues for future work.

\subsection{Converse of Kraft's Inequality for Average Lengths}

The converse of Kraft's inequality
for non-integer average lengths of indeterminate length strings is still open.
If $0<p<1$ then there is no string of average length $-\log(p)$
since superpositions of the empty string $\ket{\epsilon}$ are not prefix free.
It still remains open to find, for example, if there are three orthogonal
strings of average length $-\log_2(3)$. 

\subsection{Bounds on Known Lossless Quantum Compression}

We did not show how to bound the rate of Bostr\"{o}m and Felbinger's
lossless compression scheme \cite{bos02} by the rate of prefix free
lossless quantum compression. It is likely that the relationship can
be found by looking at the relationships between one-one coding and
prefix coding for classical codes \cite{li97,blundo,leung}.

\subsection{Applications of Lossless Quantum Compression}

Many open problems in quantum information theory \cite{nielsen,vlatkobook}, such as entanglement catalysis \cite{plenio}, are phrased
as ``Can this state be transformed into that state exactly and without error
subject to these conditions?". Maybe lossless quantum compression could be applied
to solve some of these problems. Bostr\"{o}m and Felbinger \cite{bos02} pointed out there
might also be interesting applications of variable length compression in quantum cryptography to securely
transfer data. We guess that if the base
length compression is used it minimises the probability that Eve learns any information whereas
the average length compression minimises the average amount of information that Eve learns.

\subsection{Lossy Compression of a Mixture of Mixtures}

A well-known open problem \cite{nielsen,vlatkobook} in quantum information is to find the (lossy) compression rate of a mixture
of mixtures (this problem is related to the Holevo bound). It might be simpler to find the compression rate in terms of the average length
of a variable length code \cite{sch01} rather than in terms of Schumacher coding.

\section{Acknowledgements}

We are supported by the EPSRC grant GR/S34090
and the EU Sixth Framework Programme (Project SecoQC: \textit
{Development of a Global Network for Secure Communication based on Quantum Cryptography}).
C.R. thanks Vlatko Vedral who has
supported her throughout her PhD and the writing of this paper. 

\bibliographystyle{prsty}
\bibliography{LosslessQuantumCompression}

\end{document}